\documentclass[a4paper,12pt,reqno,superscriptaddress,showkeys,nofootinbib]{revtex4}
\usepackage[centertags]{amsmath}
\usepackage{amsfonts}
\usepackage{amssymb}
\usepackage{amsthm}
\usepackage{newlfont}
\usepackage{stmaryrd}
\usepackage{mathrsfs}
\usepackage{mathtools}
\usepackage{euscript}
\usepackage{graphicx}
\usepackage{enumerate}
\usepackage[normalem]{ulem} % for strikeout text with \sout

\usepackage{color}
\definecolor{olive-green}{RGB}{60, 128, 49}

\usepackage{tikz}
\usepackage{pgf}
\usetikzlibrary{positioning,fit,calc}
\usetikzlibrary{automata,shapes}
\usetikzlibrary {arrows.meta}
\usepackage{wrapfig}
\usepackage{subfigure}
\usepackage{amscd}
\usepackage{hyperref}

\usepackage{changes}

\theoremstyle{plain}

\theoremstyle{definition}

\theoremstyle{remark}

\let\ve=\varepsilon

\newcommand{\be}{\begin{equation}}
\newcommand{\en}{\end{equation}}

\newcommand{\bbZ}{{\mathbb Z}}

\newcommand{\opunit}{\text{1}\kern-0.22em\text{l}}

\DeclareMathAlphabet{\mathpzc}{OT1}{pzc}{m}{it}

\let\oldsqrt\sqrt
\def\sqrt{\mathpalette\DHLhksqrt}
\def\DHLhksqrt#1#2{%
	\setbox0=\hbox{$#1\oldsqrt{#2\,}$}\dimen0=\ht0
	\advance\dimen0-0.2\ht0
	\setbox2=\hbox{\vrule height\ht0 depth -\dimen0}%
	{\box0\lower0.4pt\box2}}

\let\ve=\varepsilon

\let\be=\beta

\DeclareMathAlphabet{\mathpzc}{OT1}{pzc}{m}{it}

\def\bea{\begin{eqnarray}}
\def\eea{\end{eqnarray}}
\def\ba{\begin{array}}
	\def\ea{\end{array}}

\usepackage{changes}

\begin{document}

\title{Frenetic steering in a nonequilibrium graph}
\author{Bram Lefebvre and Christian Maes\\
Instituut voor Theoretische Fysica, KU Leuven, Belgium}
\email{christian.maes@kuleuven.be}%\affiliation{Gent, Belgium}
%\author{Christian Maes}
%\affiliation{Instituut voor Theoretische Fysica, KU Leuven, Belgium}

%\date{\today}

\begin{abstract}
In  traditional recognition tasks of neural networks a potential landscape or cost function guides the system towards patterns using a gradient dynamics.  That is not how the brain works as its dynamics is far from equilibrium. We present an alternative and proof of principle for pattern recovery in a nonequilibrium model whereby only time-symmetric kinetics are altered. As a mathematical model, a random walker on a randomly-oriented complete graph is subject to a finite driving in the direction of the arcs. Some vertices of the graph represent patterns. A first algorithm constructs basins of attraction for these patterns. A second algorithm updates the time-symmetric factors in the transition rates, in order for the walker to quickly reach a pattern and remain there for a sufficiently long time, whenever starting from a vertex in its basin of attraction. 
\end{abstract}

\maketitle

%\tableofcontents
\section{Introduction}
Biological activity is closely dependent on nonequilibrium conditions.  More specifically, for what is probably the most complicated system on Earth, it is rather easy to verify that the human brain functions far from equilibrium.  The evidence covers various levels of description, starting at the smallest scales of individual neurons and their nonequilibrium dynamics  \cite{sak1}. On the other end, using whole-brain imaging, large-scale nonequilibrium dynamics can be detected as well  \cite{ly}. That should not be surprising, as the nonequilibrium dynamics of biological systems are required to avoid maximum entropy states, i.e., postponing thermal death by constant energy consumption and heat dissipation, \cite{schr}.\\
In fact, the brain is one of the most demanding open systems in the body: fast synaptic signaling of neurons is  metabolically
expensive, e.g. from the transport of electric charges through pumps, \cite{kar1,kar2}. 
Even though the human brain constitutes only 2\% of the body weight, the energy-consuming processes that ensure proper brain function account for approximately
25\% of total body glucose--utilization,  taking 20\% of the body’s oxygen supply at rest \cite{al,swan}.\\
Also directly by inspecting time-series of the brain and neuronal activity, the breaking of time-reversal symmetry has been established.   In \cite{ly,deco} (using a similar method to \cite{seif}),  it was  found that brain activity time series recorded during the conscious state significantly differ from the inverted version of the same data; that was not the case for unconscious states.
It supports the idea that when the brain dynamics is chemically or physically farther from equilibrium, the recorded time series signals are irreversible. A lower level of consciousness appears to bring it closer to equilibrium and the signal also becomes more time-symmetric.  In particular, the electrocorticography signals from non-human primates under different levels of consciousness show  that temporal irreversibility of neural dynamics relates to different states of consciousness, \cite{irr1}. \\
Indeed, if the physical or cognitive tasks
become more demanding, the brain produces more entropy. Yet, interestingly, the metabolic rate is only increased by about (not more than) 10\% when the brain is stimulated, as has been demonstrated by imaging experiments  \cite{shul}. As proposed by Raichle and Mintun in \cite{rai}, the latter result suggests that the intrinsic activity of the brain, which counts as the baseline neural activity, can be as significant as, if not more than, acquiring new information and responding to changing
contingencies.  Following that line of thinking, in what follows we  fix the metabolic rate at a high level, i.e., the irreversible work done by the chemical fuel to transport activity between two ``close'' regions is considered  given.  Yet, the process of storing and memorizing patterns will involve frenetic (i.e., nondissipative and time-symmetric) components while taking place against that background of constant driving.\\

In nonequilibrium models, adding a potential to guide the system towards stored patterns does not work as used in traditional setups of neural networks and machine learning.  In other words, when time-reversibility is absent in the open system stationary dynamics, we should not expect a thermodynamic free energy landscape, cost, or loss function to guide the variational relaxation to a nearby pattern. Even the traditional methods of irreversible thermodynamics, where fluxes are linear functions of driving forces, would not do as they typically do not take into account fluctuations and nonlinearities arising from digital processing.  However, the positive news is that for nonequilibrium processes, even in the Markov approximation, time-symmetric kinetic parameters start to matter greatly; see e.g. \cite{fren,bai,nondis}. We explain more on that phenomenon in Section \ref{frst}. In fact, population selection, kinetic proofreading, error-correcting, and pattern recognition are different melodies on the same theme \cite{nondis,hopf,cur}.  They all occur in biological systems and under nonequilibrium driving, e.g. in the form of ATP hydrolysis, they make use of time-symmetric kinetics to obtain the required goal.  Those kinetic parameters are invisible in the stationary (equilibrium) population statistics  for detailed balance dynamics but they can make all the difference out of equilibrium. \\

In the present paper, we take seriously the nonequilibrium condition of neuronal networks but we do not try to model neurons and their connections directly.  Rather, we take a coarse-grained view and think of vertices in a complete graph as possibly different locations of activity.  We will have a random walker hopping to neighboring vertices to represent the transport of that activity, e.g. realized by neuronal firing and charge transport. The goal of the work is to recover patterns stored within nodes.  Indeed, a ``pattern'' is represented by a node in the graph, and the arrival of the random walker at such a pattern vertex is meant to correspond to the cognitive activity of recovery. Quite naturally, in our setup as in real brains, no direct image of a pattern is simply copied on or represented in an actual neuronal configuration.\\ 
Far from providing a neurophysiological scheme, the goal of the paper is to create a novel context of ideas for pattern storage and recovery in bioactive systems.  There are two learning steps. 
In the first step the brain creates basins of attraction for the various patterns. In the second, it learns to connect the locations in each basin with its corresponding pattern by altering the frenetic components of the dynamics of the random walk. We call the general idea of the second step frenetic steering. \\

%First, the brain creates basins of attraction for the various patterns, and second, it learns to connect the locations in each basin with \textcolor{blue}{\sout{a} its corresponding} pattern by altering\\
The next section gives more specific background about the main idea of the paper in the context of Markov jump processes.  Section \ref{section:problem} specifies the problem to be solved, which is basically about classifying states and finding the representative pattern in each class without gradient dynamics.  The first learning step, about generating basins of attraction, is explained in Section \ref{section:dec}.  Section \ref{sec:frenetic_steering} gives the method of frenetic steering.  The performance of the algorithm is discussed in Section \ref{section:performance}. 

\section{Frenesy {\it versus} entropy}\label{frst}
An important and general aspect of growth and functioning is selection. If we leave things to equilibrium, selection needs to be based on thermodynamic landscaping. We basically modify the thermodynamic potential(s) and invoke the thermodynamic variational principle(s) to select a certain macroscopic condition.  It is a version of maximizing the entropy for given constraints.  The implementation is via gradient flow in that landscape toward the selected minima. A mesoscopic version of that is obtained from the detailed balance condition of say Markov jump processes. We imagine then a finite irreducible graph, and transition rates $k(x,y)>0$ for hopping over an edge $x\rightarrow y$.  Let us use the parametrization
\begin{equation}\label{para}
k(x,y) = a(x,y)\,e^{s(x,y)/2}
\end{equation}
where $a(x,y) = a(y,x)>0$ is symmetric and $s(x,y) =- s(y,x)$ is antisymmetric.  Detailed balance means that there exists a potential $V$ defined on the vertices so that for all edges, $s(x,y) = \beta[V(x) - V(y)]$, where $\beta$ is some inverse noise strength (like inverse temperature). In that case, the symmetric prefactors $a(x,y)$ become irrelevant for the static fluctuations. In fact, the stationary distribution of that reversible Markov process is simply the Gibbs equilibrium 
\[
\rho_\text{eq}(x) = \frac 1{Z}\,e^{-\beta V(x)}
\]
concentrated on the minima of potential $V$ for large $\beta$.  That equilibrium distribution is minimizing the free energy functional
\[
\frac 1{\beta}F[\mu] = \sum_x V(x) \mu(x)  + \sum_x \mu(x)\log\mu(x) \geq \frac 1{\beta}F(\rho_\text{eq}) = - \frac 1{\beta}\,\log Z
\]
From the moment detailed balance is violated, the symmetric prefactors $a(x,y)$ get into play.  Then, they do contribute to changing the fluctuations, dynamic \'and static. The stationary distribution is no longer thermodynamically determined.  On the constructive side, we can use the $a(x,y)$ to modify and even select the stationary distribution.  That would be an application of the (idea of the) blowtorch theorem \cite{land,heatb,lowT}, where the time-symmetric kinetics (or, frenetic contribution) starts to govern the process.  For the general concept of frenesy, we refer to \cite{fren,nondis}. This then is the basic idea of the paper: use the symmetric prefactors $a(x,y)$ in \eqref{para}  to steer the system toward the selected patterns, and that becomes possible from the moment detailed balance is broken. In other words, under nonequilibrium conditions, we can change currents and occupations (i.e., steer them) by modifying time-symmetric dynamical activity; see also \cite{int}.\\

For a quick illustration, we take the simplest example.  Suppose we have a cycle with a uniform driving.  We are allowed to choose the time-symmetric part in the transition rates.  Specifically, we suppose a random walk on the ring $\bbZ_m$ with $m$ vertices and transition rates
\begin{equation}\label{ri}
k(x,x+1) = a_x\,e^{\ve/2}, \quad k(x+1,x)= a_x\,e^{-\ve/2}
\end{equation}
where $\ve \gg 1$ is the large driving.  Then, the stationary distribution is
\[
\rho^s(x) \propto \frac 1{a_{x}},\qquad x\in \bbZ_m
\]
completely determined (when $\ve \gg 1$) by the symmetric factors.  Hence, those activities $a_x$ decide the most probable state.  That is obviously impossible under detailed balance, when $\ve =0$.  Similar aspects hold then for the expected escape rates and currents.\\
More generally, in so-called Escherian systems, \cite{heatb,lowT}, any two vertices in a graph are connected both by endothermic and exothermic paths (impossible under detailed balance).  Then, the stationary distribution is completely determined by the escape rates (inverse of lifetimes).

\section{Setting the problem}
\label{section:problem}
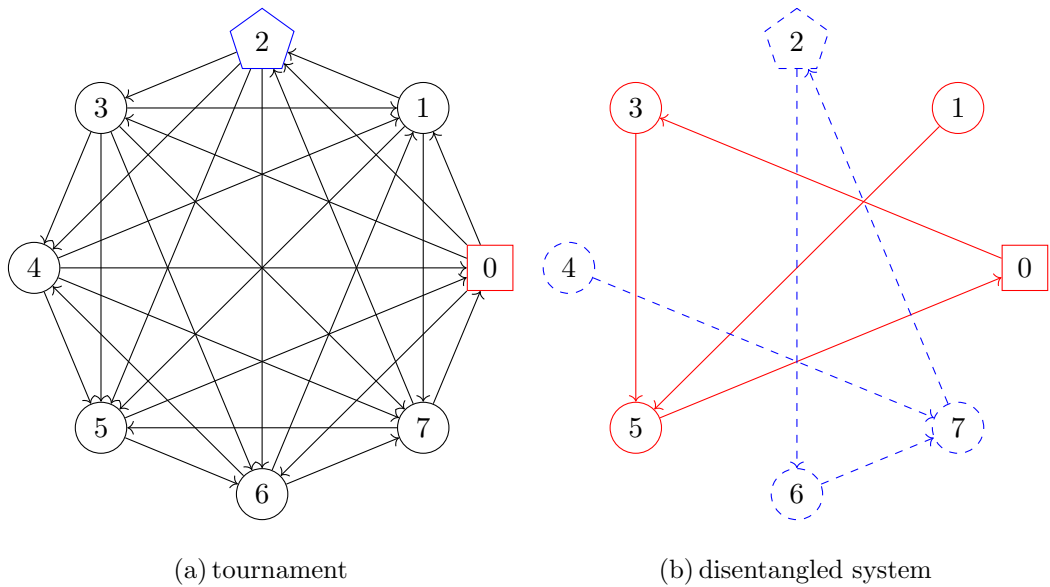
\begin{figure}
    \centering
    \subfigure[\,tournament]{
        \begin{tikzpicture}[
        vertex/.style={circle, draw, minimum size=5mm}
        ]
    
        \draw node[minimum size=6mm, draw=red](0) at (3, 0) {0};
        \draw node[vertex](1) at (2.12, 2.12) {1};
        \draw node[regular polygon, regular polygon sides=5, minimum size=5mm, draw=blue](2) at (0, 3) {2};
        \draw node[vertex](3) at (-2.12, 2.12) {3};
        \draw node[vertex](4) at (-3, 0) {4};
        \draw node[vertex](5) at (-2.12, -2.12) {5};
        \draw node[vertex](6) at (0, -3) {6};
        \draw node[vertex](7) at (2.12, -2.12) {7};
        
        \draw [->] (0) -- (1);
        \draw [->] (0) -- (2);
        \draw [->] (0) -- (3);
        \draw [->] (0) -- (6);
        \draw [->] (1) -- (2);
        \draw [->] (1) -- (5);
        \draw [->] (1) -- (7);
        \draw [->] (2) -- (3);
        \draw [->] (2) -- (4);
        \draw [->] (2) -- (5);
        \draw [->] (2) -- (6);
        \draw [->] (3) -- (1);
        \draw [->] (3) -- (4);
        \draw [->] (3) -- (5);
        \draw [->] (3) -- (6);
        \draw [->] (3) -- (7);
        \draw [->] (4) -- (0);
        \draw [->] (4) -- (1);
        \draw [->] (4) -- (5);
        \draw [->] (4) -- (7);
        \draw [->] (5) -- (0);
        \draw [->] (5) -- (6);
        \draw [->] (6) -- (1);
        \draw [->] (6) -- (4);
        \draw [->] (6) -- (7);
        \draw [->] (7) -- (0);
        \draw [->] (7) -- (2);
        \draw [->] (7) -- (5);
    
        \end{tikzpicture}
    
        \label{fig:graph}
        %Example of a tournament with patterns $x_1=0$ (red square) and  $x_2=2$ (blue pentagon).
    }
    \subfigure[\,disentangled system]{
        \begin{tikzpicture}[
        vertex/.style={circle, draw, minimum size=5mm},
        basin_0_vertex/.style={circle, draw=red, minimum size=5mm},
        basin_1_vertex/.style={dashed, circle, draw=blue, minimum size=5mm},
        basin_0_arc/.style={red, ->},
        basin_1_arc/.style={dashed, blue, ->},
        ]
        
        \draw node[minimum size=6mm, draw=red](0) at (3, 0) {0};
        \draw node[basin_0_vertex](1) at (2.12, 2.12) {1};
        \draw node[dashed, regular polygon, regular polygon sides=5, minimum size=5mm, draw=blue](2) at (0, 3) {2};
        \draw node[basin_0_vertex](3) at (-2.12, 2.12) {3};
        \draw node[basin_1_vertex](4) at (-3, 0) {4};
        \draw node[basin_0_vertex](5) at (-2.12, -2.12) {5};
        \draw node[basin_1_vertex](6) at (0, -3) {6};
        \draw node[basin_1_vertex](7) at (2.12, -2.12) {7};
        
        \draw [basin_0_arc] (0) -- (3);
        \draw [basin_0_arc] (3) -- (5);
        \draw [basin_0_arc] (5) -- (0);
        \draw [basin_0_arc] (1) -- (5);
        
        \draw [basin_1_arc] (2) -- (6);
        \draw [basin_1_arc] (6) -- (7);
        \draw [basin_1_arc] (7) -- (2);
        \draw [basin_1_arc] (4) -- (7);
        
        \end{tikzpicture}
        \label{fig:disentangled_graph}
        % Example of an oriented graph obtained by executing the disentanglement algorithm on the tournament in Fig. \ref{fig:graph}. The patterns are $x_1=0$ (red square) and  $x_2=2$ (blue pentagon) and the basins are $B_1=\{0, 1, 3, 5\}$ (red, full line) and $B_2=\{2, 4, 6, 7\}$ (blue, dashed line).
        
    }
    \caption{Example of a tournament (a) and an oriented graph (b), obtained by executing the disentanglement algorithm on that tournament. The patterns are $x_1=0$ (red square) and  $x_2=2$ (blue pentagon) and the basins are $B_1=\{0, 1, 3, 5\}$ (red, full line) and $B_2=\{2, 4, 6, 7\}$ (blue, dashed line).}

\end{figure}

We make the problem specific by choosing the complete graph with $N$ vertices. We think of it as the network in the background.  That is a simplification for exploring the ideas and at the same time, a complete graph is the most symmetric point of departure.  We randomly orient every edge $\{x,y\}$ by assigning, independently, variables $\sigma(x,y) =  -\sigma(y,x) \in \{-1,+1\}$ to its two orientations $(x,y): x\rightarrow y$ and $(y,x): y\rightarrow x$. When $\sigma(x,y)=+1$, we say that the orientation $x\rightarrow y$ is preferred. 
In that way, we may visualize the preferred direction by arrows assigned to every edge of the complete graph, making it into a tournament. Fig.~\ref{fig:graph} shows an example of a tournament. \\

At the same time, we put $s(x,y) =\ve\, \sigma(x,y), s(y,x) = -s(x,y)$ with $\ve>0$ giving the magnitude of a nonequilibrium driving. Many loops now emerge in the network over which a current can be maintained. That random background remains unaltered during all of the following constructions and learning. In addition, we attach an unoriented weight $a(x,y)=a(y,x) \geq 0$ to each edge, which, in contrast, will be altered during learning following a version of the Hebb rule \cite{heb}. They realize synaptic strengths on a coarse-grained level.\\

%trained to learn to find the pattern.  We call it {\it frenetic steering} in a fixed dissipative background. The weights are time-symmetric activity parameters.\\
Next, we define a random walker $X_t$ on the graph.  It is interpreted as transmitter of firing activity. 
The walker has transition rates $k(x,y) \geq 0$ to hop over the edge $x\rightarrow y$ with
\begin{equation}
    \sqrt{k(x,y)k(y,x)} = a(x,y), \qquad     \log\frac{k(x,y)}{k(y,x)} = s(x,y)
\end{equation}
The weights $\{a(x,y)\}$ are the time-symmetric activity parameters of the random walker.  They are variable and will be updated to steer the random walker.\\ 
The goal for the walker is to reach a pattern, interpreted as a condition of cognitive recall.  More precisely, we randomly select $k>1$ patterns, i.e. $k$ different vertices $\{x_1,x_2,\ldots,x_k\}$ on the graph, which serve as targets for this random walker. 
The problem is to train the walker by adapting its time-symmetric activity parameters so that the following holds:\\
%for ``most'' orientations $\sigma$ of the graph.  Fixing times $T>0$ and $\tau > 0$ (to be specified below): \\
- Draw a vertex $u$ and start the random walker from $X_0=u$.  We want that there is a pattern $x_i$ so that with high probability $X_T = x_i$ at time $T$ and that the residence time-interval 
$I=[t_i,t_f]\ni T$ for which $X_t= x_i, \forall t \in I$,
 has a duration $|t_f-t_i| \geq \tau$ at least $\tau$.
We then say that $u$ is in the basin of attraction $B_i$ of pattern $x_i$.\\
- We want that for each pattern $x_i,\, i=1, \ldots, k$, the number of vertices $|B_i|$ in its basin scales like $N/k$.\\ 
Moreover, altering the time-symmetric activity parameters $\{a(x,y)\}$ will be done iteratively, via supervised learning, whereby the initial value is taken uniform: $a_\text{ini}(x, y)=a$ for some $a>0$ for all edges in the basins.\\

The first problem is to divide the network into different regions  (``colors'')  belonging to one particular pattern. Depending on the random orientations, the network stores the patterns in different basins. 

In that first learning step, we run an algorithm for the successful decomposition of the oriented graph into basins of attraction $B_i$, given any selection of patterns and for ``any'' (random) realization of the orientation of the  oriented  graph. We call it the disentanglement step.
We start with that in the next section.\\
The second (supervised) learning step towards recovery (in Section \ref{sec:frenetic_steering}) wants to achieve that starting the random walker from a vertex in $B_i$, it arrives `fast enough' (after at most a time $T$) in $x_i$ and stays there for a good while (at least a time $\tau$). We call it the frenetic steering step. One should imagine here that when the walker arrives at the pattern, it ``lights up'' announcing the ``color'' of the vertex in which the walker started.  In that interpretation, upon stimulating a region of the network  (start of the random walk), after a short time, we find out to what pattern it belongs.\\

The algorithms are implemented in python \cite{python}, where the library numpy \cite{numpy} was used for general calculations and the library matplotlib \cite{matplotlib} enabled to investigate the performance of the algorithms. 
The source code of the implementations of the algorithms can be found at \cite{git}.\\
In Section \ref{section:performance} we run the algorithms and check their performance, also as a function of $N, k, \varepsilon$ and (what will be called) the training set size.  Finally, in the conclusion Section \ref{con}, we give an outlook including more interpretation.

\section{Disentangling the basins of attraction}
\label{section:dec}
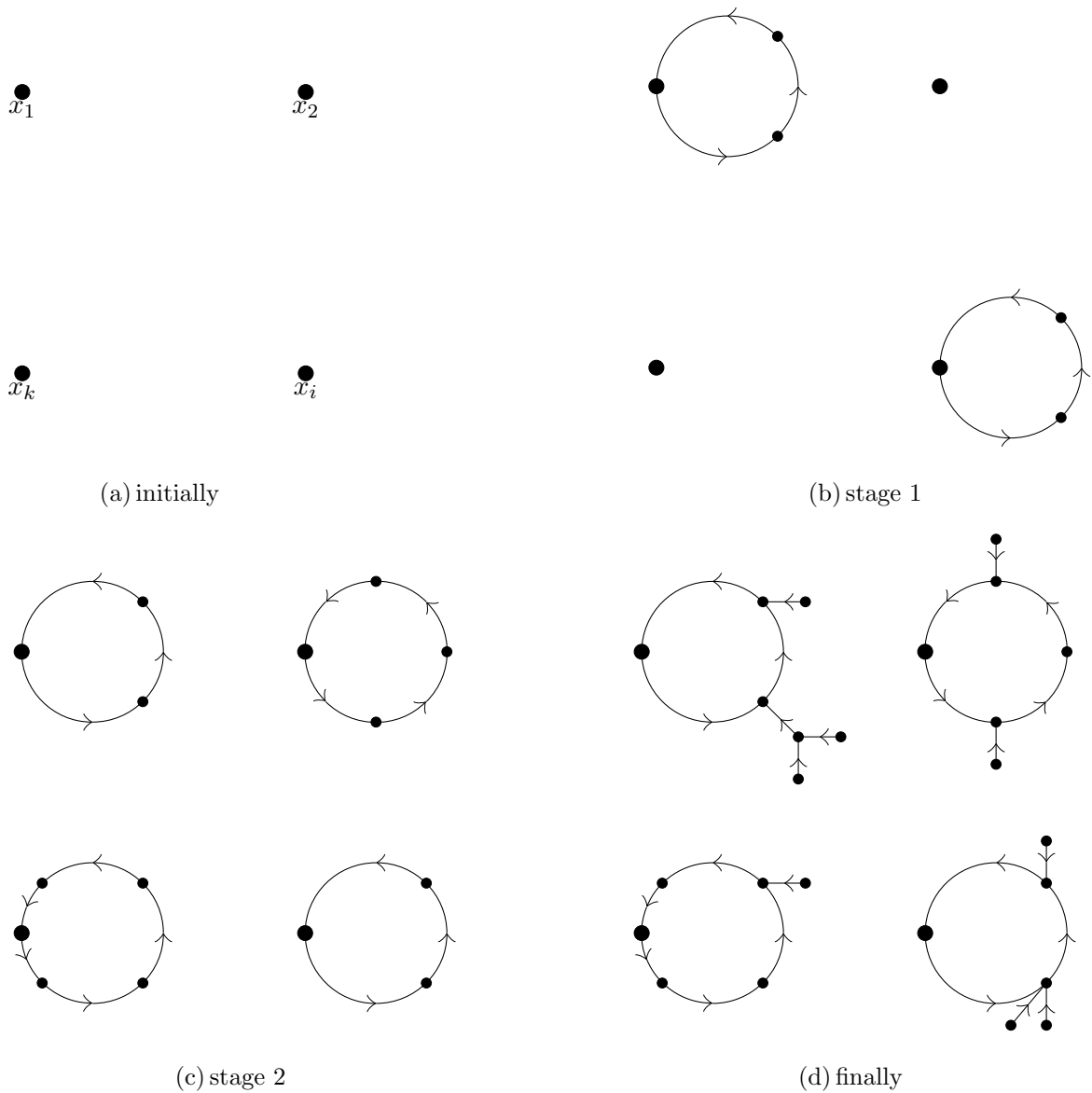
\begin{figure}
    \centering
    \subfigure[\,initially ]{
    	\begin{tikzpicture}
		\filldraw (1, 2) circle (3pt) node[align=left,   below] {$x_k$};
		\filldraw (1, 6) circle (3pt) node[align=left,   below] {$x_1$};
		\filldraw (5, 2) circle (3pt) node[align=left,   below] {$x_i$};
		\filldraw (5, 6) circle (3pt) node[align=left,   below] {$x_2$};
		\filldraw[draw=white, fill=white] (1, 1) circle (1pt);
    	
    	\end{tikzpicture}
	}
	\hspace{4cm}
    \subfigure[\,stage 1 ]{
    	\begin{tikzpicture}
		\filldraw (1, 2) circle (3pt);
		\filldraw (1, 6) circle (3pt);
		\filldraw (5, 2) circle (3pt);
		\filldraw (5, 6) circle (3pt);
		
		\filldraw (2.71, 6.71) circle (2pt);
		\filldraw (2.71, 5.29) circle (2pt);
		\draw [-{Classical TikZ Rightarrow[scale=2]}] (2, 7) arc (90:270:1);
		\draw [-{Classical TikZ Rightarrow[scale=2]}] (2, 5) arc (270:360:1);
		\draw [-{Classical TikZ Rightarrow[scale=2]}](3, 6) arc (0:90:1);

		\filldraw (6.71, 1.29) circle (2pt);
		\filldraw (6.71, 2.71) circle (2pt);
		\draw [-{Classical TikZ Rightarrow[scale=2]}] (6, 3) arc (90:270:1);
		\draw [-{Classical TikZ Rightarrow[scale=2]}] (6, 1) arc (270:360:1);
		\draw [-{Classical TikZ Rightarrow[scale=2]}] (7, 2) arc (0:90:1);

    	\end{tikzpicture}
    }

    \subfigure[\,stage 2 ]{
    	\begin{tikzpicture}
		\filldraw (1, 2) circle (3pt);
		\filldraw (1, 6) circle (3pt);
		\filldraw (5, 2) circle (3pt);
		\filldraw (5, 6) circle (3pt);
		
		\filldraw (2.71, 6.71) circle (2pt);
		\filldraw (2.71, 5.29) circle (2pt);
		\draw [-{Classical TikZ Rightarrow[scale=2]}] (2, 7) arc (90:270:1);
		\draw [-{Classical TikZ Rightarrow[scale=2]}] (2, 5) arc (270:360:1);
		\draw [-{Classical TikZ Rightarrow[scale=2]}](3, 6) arc (0:90:1);

		\filldraw (6.71, 1.29) circle (2pt);
		\filldraw (6.71, 2.71) circle (2pt);
		\draw [-{Classical TikZ Rightarrow[scale=2]}] (6, 3) arc (90:270:1);
		\draw [-{Classical TikZ Rightarrow[scale=2]}] (6, 1) arc (270:360:1);
		\draw [-{Classical TikZ Rightarrow[scale=2]}] (7, 2) arc (0:90:1);

		\filldraw (6, 5) circle (2pt);
		\filldraw (7, 6) circle (2pt);
		\filldraw (6, 7) circle (2pt);
		\draw [-{Classical TikZ Rightarrow[scale=2]}] (5.29, 6.71) arc (135:225:1);
		\draw [-{Classical TikZ Rightarrow[scale=2]}] (5.29, 5.29) arc (225:315:1);
		\draw [-{Classical TikZ Rightarrow[scale=2]}] (6.71, 5.29) arc (315:405:1);
		\draw [-{Classical TikZ Rightarrow[scale=2]}] (6.71, 6.71) arc (45:135:1);

		\filldraw (1.29, 1.29) circle (2pt);
		\filldraw (2.71, 1.29) circle (2pt);
		\filldraw (2.71, 2.71) circle (2pt);
		\filldraw (1.29, 2.71) circle (2pt);
		\draw [-{Classical TikZ Rightarrow[scale=2]}] (1.07, 2.37) arc (158:202:1);
		\draw [-{Classical TikZ Rightarrow[scale=2]}] (1.07, 1.63) arc (202:270:1);
		\draw [-{Classical TikZ Rightarrow[scale=2]}] (2, 1) arc (270:360:1);
		\draw [-{Classical TikZ Rightarrow[scale=2]}] (3, 2) arc (0:90:1);
		\draw [-{Classical TikZ Rightarrow[scale=2]}] (2, 3) arc (90:158:1);
    	
    	\filldraw[draw=white, fill=white] (6.71, 0.69) circle (2pt);
    	\end{tikzpicture}
    }
    \hspace{2cm}
    \subfigure[\,finally ]{
    	\begin{tikzpicture}
		\filldraw (1, 2) circle (3pt);
		\filldraw (1, 6) circle (3pt);
		\filldraw (5, 2) circle (3pt);
		\filldraw (5, 6) circle (3pt);
		
		\filldraw (2.71, 6.71) circle (2pt);
		\filldraw (2.71, 5.29) circle (2pt);
		\draw [-{Classical TikZ Rightarrow[scale=2]}] (2, 7) arc (90:270:1);
		\draw [-{Classical TikZ Rightarrow[scale=2]}] (2, 5) arc (270:360:1);
		\draw [-{Classical TikZ Rightarrow[scale=2]}](3, 6) arc (0:90:1);

		\filldraw (6.71, 1.29) circle (2pt);
		\filldraw (6.71, 2.71) circle (2pt);
		\draw [-{Classical TikZ Rightarrow[scale=2]}] (6, 3) arc (90:270:1);
		\draw [-{Classical TikZ Rightarrow[scale=2]}] (6, 1) arc (270:360:1);
		\draw [-{Classical TikZ Rightarrow[scale=2]}] (7, 2) arc (0:90:1);

		\filldraw (6, 5) circle (2pt);
		\filldraw (7, 6) circle (2pt);
		\filldraw (6, 7) circle (2pt);
		\draw [-{Classical TikZ Rightarrow[scale=2]}] (5.29, 6.71) arc (135:225:1);
		\draw [-{Classical TikZ Rightarrow[scale=2]}] (5.29, 5.29) arc (225:315:1);
		\draw [-{Classical TikZ Rightarrow[scale=2]}] (6.71, 5.29) arc (315:405:1);
		\draw [-{Classical TikZ Rightarrow[scale=2]}] (6.71, 6.71) arc (45:135:1);

		\filldraw (1.29, 1.29) circle (2pt);
		\filldraw (2.71, 1.29) circle (2pt);
		\filldraw (2.71, 2.71) circle (2pt);
		\filldraw (1.29, 2.71) circle (2pt);
		\draw [-{Classical TikZ Rightarrow[scale=2]}] (1.07, 2.37) arc (158:202:1);
		\draw [-{Classical TikZ Rightarrow[scale=2]}] (1.07, 1.63) arc (202:270:1);
		\draw [-{Classical TikZ Rightarrow[scale=2]}] (2, 1) arc (270:360:1);
		\draw [-{Classical TikZ Rightarrow[scale=2]}] (3, 2) arc (0:90:1);
		\draw [-{Classical TikZ Rightarrow[scale=2]}] (2, 3) arc (90:158:1);

		\filldraw (3.21, 4.79) circle (2pt);
		\draw [-{Classical TikZ Rightarrow[scale=2]}] (3.21, 4.79) -- (2.96, 5.04);
		\draw (2.96, 5.04) -- (2.71, 5.29);
		\filldraw (3.21, 4.19) circle (2pt);
		\draw [-{Classical TikZ Rightarrow[scale=2]}] (3.21, 4.19) -- (3.21, 4.49);
		\draw (3.21, 4.49) -- (3.21, 4.79);
		\filldraw (3.81, 4.79) circle (2pt);
		\draw [-{Classical TikZ Rightarrow[scale=2]}] (3.81, 4.79) -- (3.51, 4.79);
		\draw (3.51, 4.79) -- (3.21, 4.79);
		\filldraw (3.31, 6.71) circle (2pt);
		\draw [-{Classical TikZ Rightarrow[scale=2]}] (3.31, 6.71) -- (3.01, 6.71);
		\draw (3.01, 6.71) -- (2.71, 6.71);
		
		\filldraw (6, 4.4) circle (2pt);
		\draw [-{Classical TikZ Rightarrow[scale=2]}] (6, 4.4) -- (6, 4.7);
		\draw (6, 4.7) -- (6, 5);
		\filldraw (6, 7.6) circle (2pt);
		\draw [-{Classical TikZ Rightarrow[scale=2]}] (6, 7.6) -- (6, 7.3);
		\draw (6, 7.3) -- (6, 7);
		
		\filldraw (6.71, 0.69) circle (2pt);
		\draw [-{Classical TikZ Rightarrow[scale=2]}] (6.71, 0.69) -- (6.71, 0.99);
		\draw (6.71, 0.99) -- (6.71, 1.29);
		\filldraw (6.21, 0.69) circle (2pt);
		\draw [-{Classical TikZ Rightarrow[scale=2]}] (6.21, 0.69) -- (6.46, 0.99);
		\draw (6.46, 0.99) -- (6.71, 1.29);
		\filldraw (6.71, 3.31) circle (2pt);
		\draw [-{Classical TikZ Rightarrow[scale=2]}] (6.71, 3.31) -- (6.71, 3.01);
		\draw (6.71, 3.01) -- (6.71, 2.71);
		\filldraw (3.31, 2.71) circle (2pt);
		\draw [-{Classical TikZ Rightarrow[scale=2]}] (3.31, 2.71) -- (3.01, 2.71);
		\draw (3.01, 2.71) -- (2.71, 2.71);

    	\end{tikzpicture}
    }
    \caption{Cartoon indicating the stages (a)--(b)--(c)--(d) in the disentangling algorithm for 4 patterns.  First, minimal cycles are added around the pattern, after which trees are attached obtaining 4 tree-loops.}
    \label{fig:stages}
\end{figure}

In the present section, we work with a general finite and connected randomly-oriented graph. We refer to \cite{b-j, douglas} for standard results in graph theory. The goal of the disentanglement is to decompose that randomly-oriented graph into mutually disconnected subgraphs (basins of attraction).  In reality, that should also be modeled in more detailed correspondence with what happens in  ``structuring'' the (young) brain.  That is an interesting and deep problem, where also the very notion of pattern should be made compatible with the actual brain functioning of selection, recovery or recognition.  We skip that problem here and instead {\it start} from the presence of patterns to have an algorithm assigning vertices to a basin in which the edges are directed to the associated pattern.  The stages in the algorithm are illustrated in Fig.~\ref{fig:stages}, where each pattern is the hub and attractor of a tree-loop oriented toward the pattern.\\

We give a brief explanation of the algorithm represented in Fig.~\ref{fig:stages}, where the oriented graph (background network) is not indicated however.  We order the selected patterns $x_1,\dots,x_k$ and we iteratively grow a tree-loop around each pattern, i.e.,  a cycle (or oriented loop) on which we find the pattern and to which are attached (oriented) trees (or hairs, toward the cycle). The final basin $B_i$ is also a tree-loop and the algorithm ends when no more arcs can be added.\\
To begin, we look at $x_1$ in the oriented graph and see to find (with trial and error) a cycle of length $3$, i.e., with two other distinct vertices connected by an arc making the cycle $x_{11} = x_1 \rightarrow x_{12} \rightarrow x_{13} \rightarrow x_1$, with of course $x_{12}$ and $x_{13}$ also different from all other patterns.   Even if we do not find such a cycle for the first pattern, we move to pattern $x_2$ and we do the same thing: finding (or not) a cycle of length three containing $x_2$, mutually disjoint from all other previously selected vertices.  We repeat until we finish with $x_k$ and see if we can make it part of a three-cycle or not.  We have then completed stage 1; see Fig.~\ref{fig:stages}.\\  We next go again through all the patterns which have not been made part of a three-cycle and try to see if they can be part of a four-cycle, etc until each pattern is part of some (minimal) cycle (possibly none for some patterns).  We have then completed stage 2 in Fig.~\ref{fig:stages}.\\
The algorithm gets finished by adding trees (edge per edge) oriented toward one of the non-pattern vertices in the obtained cycles.  That is done in an equilibrated way to make sure that the basins all contain about the same number of vertices.  In the end, every basin is either a single vertex (no cycle was found to which the pattern belongs), or a tree-loop.   The code is freely available from github \cite{git}.\\ 

For a simple example, we turn to   Fig.~\ref{fig:graph} showing a tournament with patterns $x_1=0$ (red square) and  $x_2=2$ (blue pentagon).  Fig.~\ref{fig:disentangled_graph} shows the disentangled system obtained by executing the disentanglement algorithm.  We obtain for basin $B_1$ the cycle $(0, 3, 5, 0)$ and the tree is $(1,5)$, and for basin $B_2$ the cycle is $(2, 6, 7, 2)$ and the tree is $(4,7)$. \\
Let us reconstruct what has happened.
We first look for a cycle for each basin. We start with basin $B_1$ and take the subgraph by removing all vertices that are elements of the other basins. The only other basin is $B_2$ and it only contains the vertex $2$, so we end up with the vertices $\{0, 1, 3, 4, 5, 6, 7\}$. We search for a cycle of length 3 and we find $(0, 3, 5, 0)$. These vertices and the arcs are added to $B_1$. Then we go to the other basin $B_2$. The subgraph we have to look at has the vertices $\{1, 2, 4, 6, 7\}$ because $B_1=\{0, 3, 5\}$. We search for a cycle of length 3 and find  $(2, 6, 7, 2)$ and add these vertices and the arcs to $B_2$. We go back to $B_1$. The only vertices that are not assigned to any basin are $1$ and $4$. We take $(1,5)$ as tree. We go to basin $B_2$. The only vertex left is $4$. We add the tree $(4,7)$.\\

We can view the disentangling algorithm as making decisions about which edges in the original oriented graph to keep and which to remove.
However, the disentangling is not merely a clustering or coloring of the graph as we wish to have each basin having all edges oriented toward the corresponding pattern. In other words, the leading idea of disentangling is to provide {\it fast lanes} to each pattern. Indeed, with reference to the next section of frenetically steering a random walker, the purpose  is to set $a(x,y)=0$ for the unselected edges $\{x,y\}$ and to remain with subgraphs (basins) where in each basin the orientations point to the pattern. 
Yet, on the basins, the arcs only represent the preferred direction of the random walker while jumps against the direction of the arcs remain possible.\\
The performance of the algorithm is discussed in Section \ref{perd}.
Clearly, it is not surprising that such a disentanglement algorithm exists and works well. The main purpose is to provide proof of principle for the idea of using frenetic steering to recover a pattern, which is next.

\section{Frenetic steering}\label{sec:frenetic_steering}
We come to the second learning step, starting after the disentangling.  Indeed, after the disentanglement we have divided the graph into ``rooms with a different color'', each belonging to a unique pattern, wherein corridors give a preferred orientation as borrowed from the background network. We put $a(x,y)=0$ for all edges that are removed in the disentanglement step.\\
When starting anywhere in one of those rooms, we want to quickly recall the corresponding pattern.  In order to achieve that, we alter the activity parameters during a learning process; they will no longer be uniform. Note that we do not want to inspect the basin of attraction globally and decide on a specific choice of time-symmetric activities in each case of disentangling.  Rather, we give an algorithm that always uses the same updating rules to teach a random walker locally and iteratively how to quickly reach the pattern and stay there long enough.\\

The general idea is that the transition rates leaving a pattern be small and that for each state in a basin, there should be (a) fast path(s) `toward' the corresponding pattern. Supervised learning is to generate paths of the random walk and, if a path is not desirable, to change the activity parameters in a strategic Hebbian way. We call it frenetic steering because the updating only concerns those unoriented (time-symmetric) parameters.\\
In that teaching,  paths are generated by randomly picking an initial state and letting the system evolve until it leaves the state it was in at time $T$. The number of times we do this defines the {\it training set size} $n$. \\
A successful path has the walker arriving in the pattern in a time less than or equal to $T$, to not leave before the time $T$, and to stay longer than the minimal residence time $\tau$.\\ 
That leads to three ways in which a path can be unsuccessful. \\

(1) the walker has left the pattern before time $T$:\\
We decrease the activity parameters for all the transitions in the path leaving the pattern which are in the direction of the driving.\\ 

(2) the walker never visited the pattern before time $T$:\\
For every state in the path, we look at the next state in the path. If this next state in the path is in the direction of the driving, we increase the activity parameters for the transition from the original state to this next state. If the next state in the path is against the direction of the driving, we increase the activity parameters for all the transitions leaving the original state which are in the direction of the driving.  For the last state in the path we also increase the activity parameters for all the transitions leaving this state which are in the direction of the driving. We take care to not increase the same activity parameters multiple times during the handling of this case.\\

(3) the walker has never left the pattern before time $T$ and was present in the pattern at time $T$ but the actual residence time was smaller than the minimal residence time $\tau$:\\
We decrease the activity parameters for all transitions in the direction of the driving from the last state the system is in, to a state outside the pattern.\\

In the iteration, the activity parameters change value proportional to their current value. Introducing a learning rate $R$, where $0\leq R\leq 1$, we take the new or updated symmetric part as
\begin{equation}\label{rr}
    a_\text{new}(x,y) = \left\{
    \begin{array}{rl}
    \frac 1{R}\,a_\text{old}(x,y) & \text{when increasing }\\
    R\,a_\text{old}(x,y) & \text{when decreasing }
    \end{array} \right.
\end{equation}

The input parameters for the entire algorithm are 
\begin{itemize}
    \item the travel time $T$ and residence time $\tau$,
    \item driving value $\varepsilon$ and initial value for the time-symmetric activity parameters $a$,
    \item learning rate $R$, where $0<R<1$, and
    \end{itemize}
the success will depend on the training set size $n$, the number $N$ of vertices, and the pattern number $k$.\\
As the initial uniform value (at the beginning of the frenetic steering) for the activity parameters $a_\text{ini}(x,y) = a$ (for the edges that have not been removed by the disentanglement step), we take
\begin{equation}
    a = \frac{20}{T}\exp\Big(\frac{-\varepsilon}{2}\Big)
    \label{eq:a}
\end{equation}
The prefactor (here equal to $20$) is taken large enough to allow for sufficient dynamical activity to start from. Then, most of the paths generated during the learning are of type (1), which is the situation in which the algorithm performs best. There is however a large range of values for $a$ for which the algorithm performs well but for small values, say $a<1$, the walker can more easily get trapped outside the pattern and for very large values the performance might also drop, but we have not observed this.\\
In what follows we take the values $T=1$, $\varepsilon = 5$, $R=0.5$, $\tau=0.2$. The choice for $R=0.5$ seemed the most balanced, but the algorithm works well for many values for $R$. It works even better for smaller values for $R$, but the performance starts to decrease for bigger values for $R$.\\

Of course, success is never guaranteed. There is the danger that the walker gets trapped outside the pattern. Remember also that the arcs only represent the preferred direction of the random walker and that jumps against the direction of the arcs are possible. At any rate, the algorithm provides proof of principle to demonstrate that frenetic steering is capable of pattern recovery. The performance of the algorithm is discussed in Section \ref{perf}, i.e. mostly how the success depends on $n$ (the training set size), and $k$ (the number of patterns).\\

\begin{figure}[ht]
    \centering
    \subfigure[\,before learning]{
        \includegraphics[height=0.25\textheight]{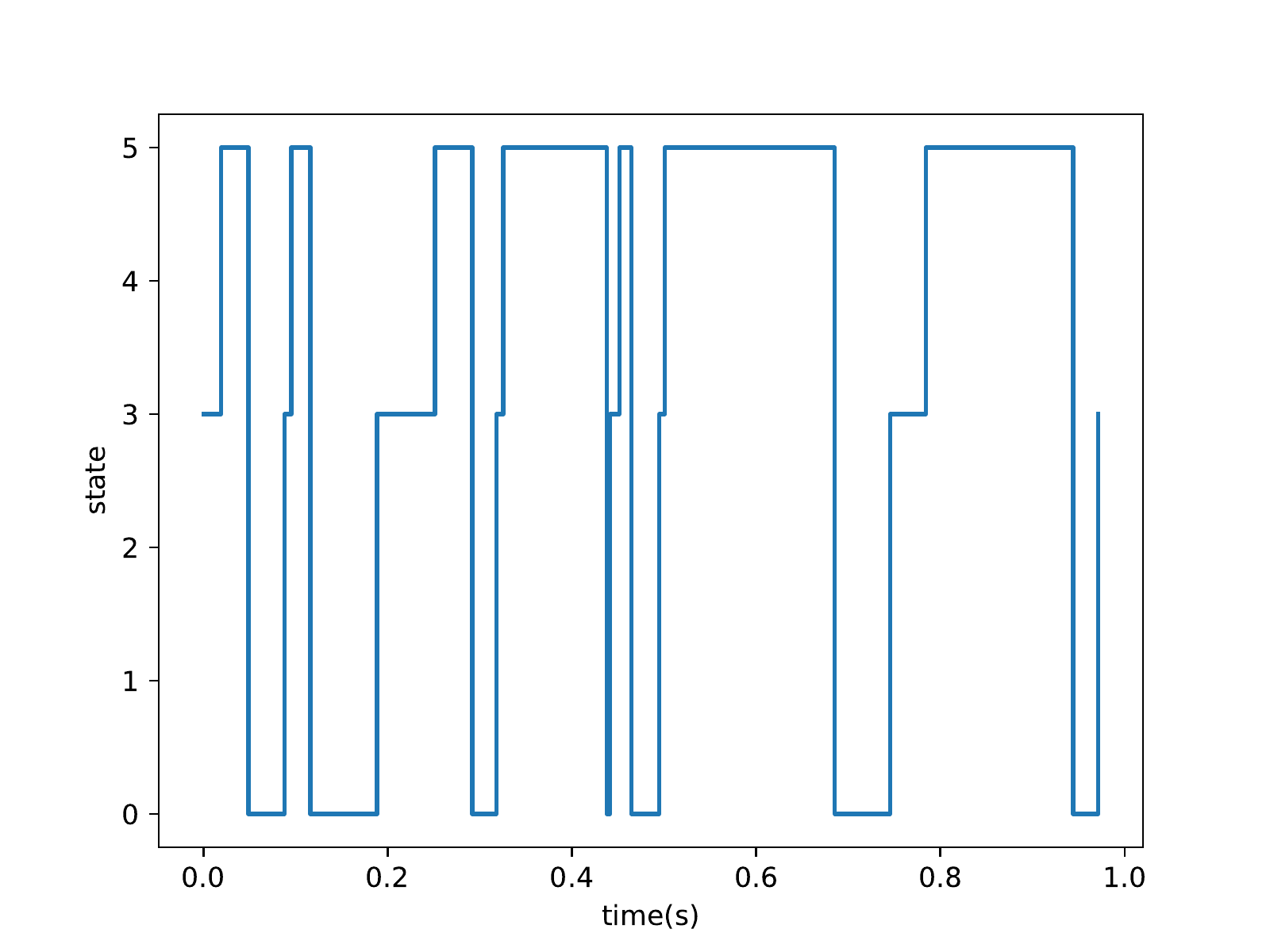}
        \label{fig:non_trained_path}
    }
    \subfigure[\,after learning]{
        \includegraphics[height=0.25\textheight]{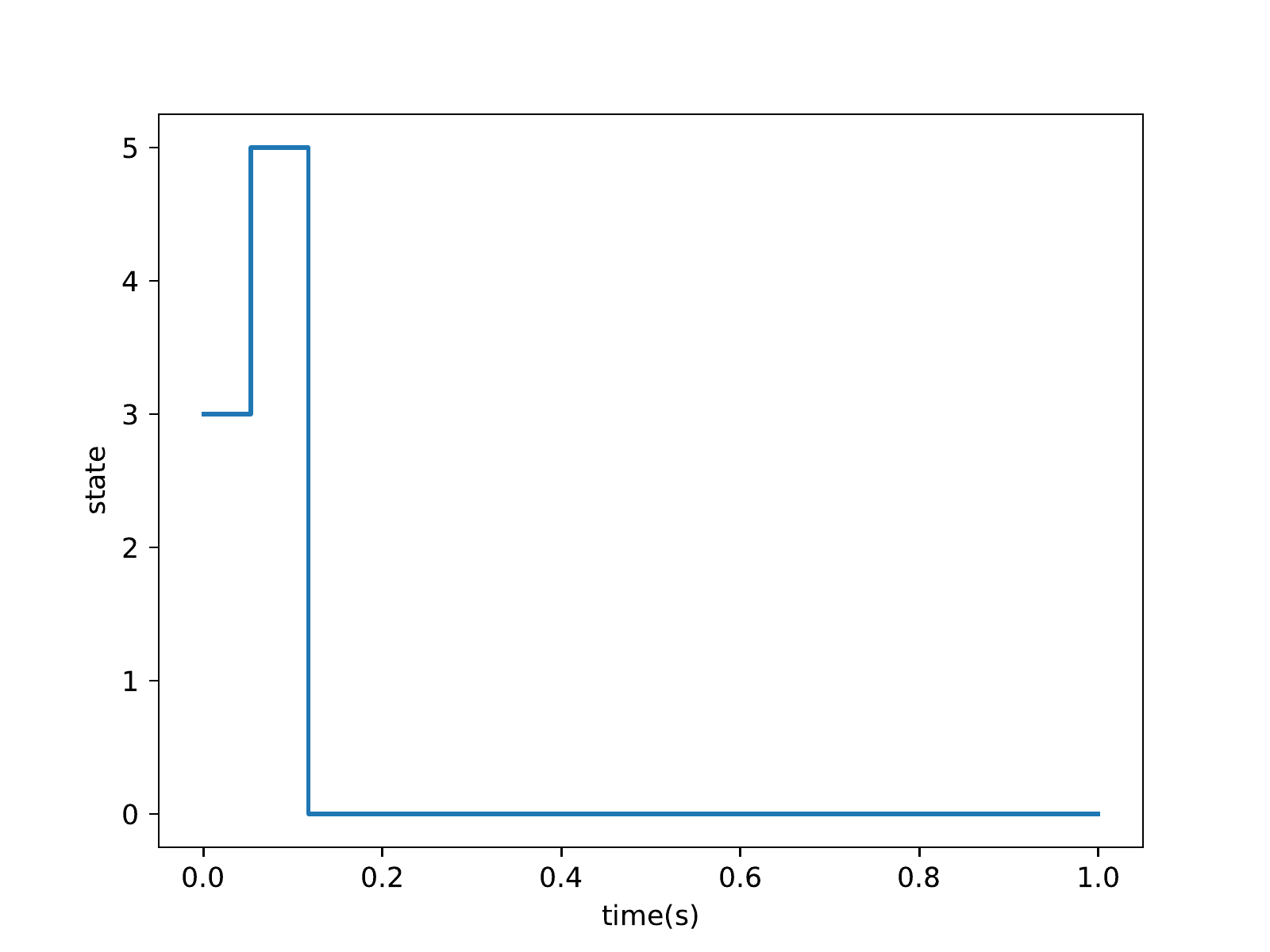}
        \label{fig:trained_path}
    }
    \caption{`Typical' path for a random walker on the disentangled system shown in Fig. \ref{fig:disentangled_graph} both before and after the learning.}

\end{figure}

Let us illustrate the algorithm by considering paths on the oriented graph of Fig.~\ref{fig:disentangled_graph}.  Remember that the paths are observed until the system leaves the state it is in at time $T$.\\
The path $(5, 0, 3)$  leaves a pattern so that we are in situation (1), and we, therefore, decrease the value of the activity parameter $a(0, 3)$. The path $(3, 5)$ never visits a pattern so we are in situation (2), and we, therefore, increase the value of the activity parameters $a(3, 5)$ and $a(5, 0)$. 
% Consider $(3, 5, 3)$ as generated path. It never visits a pattern state so we are in situation (2), and we, therefore, increase the value of the activity parameters $a(3, 7)$, $a(7, 0)$ and $a(7, 5)$. 
The path $(3, 5, 0)$ has a residence time smaller than the minimal residence time. It never leaves the pattern before time $T$ and is present in the pattern at time $T$ but the actual residence time is smaller than the desired residence time.  We are therefore in situation (3) and we decrease the activity parameter $a(0, 3)$.\\ 
Looking at Figs.~\ref{fig:non_trained_path}--\ref{fig:trained_path},  we take 3 as the initial state, which is an element of basin $B_1$.  Then, a `typical' path before `learning' would look like the one in Fig.~\ref{fig:non_trained_path}: the system does not really care to stay in the pattern $x_1=0$. When the learning step is performed, a `typical' path looks like  Fig.~\ref{fig:trained_path}, and the system stays in the pattern once it has arrived there.

\section{Performance}\label{section:performance}

\subsection{Performance of disentangling}\label{perd}
As we start the performance analysis for complete graphs, for reasons of symmetry, we like the basins emerging after disentangling to be of similar size.  To check, we generate 100 tournaments (orientations of the complete graph) and for each of these tournaments, we run the algorithm described in Section \ref{section:dec}. We thus obtain 100 disentangled systems. The size of a basin $|B|$ is the number of its vertices, not including the pattern vertex, and with the applied algorithm we reach an average size equal to $N/k - 1$. \\

%As a first example, we take $N=50, k=5$. The average size of a basin is 9 in this case. For the 100 disentangled systems we generated all of them had a size of 9.

%re were 2 for which the size of the smallest basin was 5, 23 for which the size of the smallest basin was 6, 43 for which the size of the smallest basin was 7, 31 for which the size of the smallest basin was 8 and 1 for which the size of the smallest basin was 9.\\

\begin{figure}
    \centering
    \includegraphics[height=0.3\textheight]{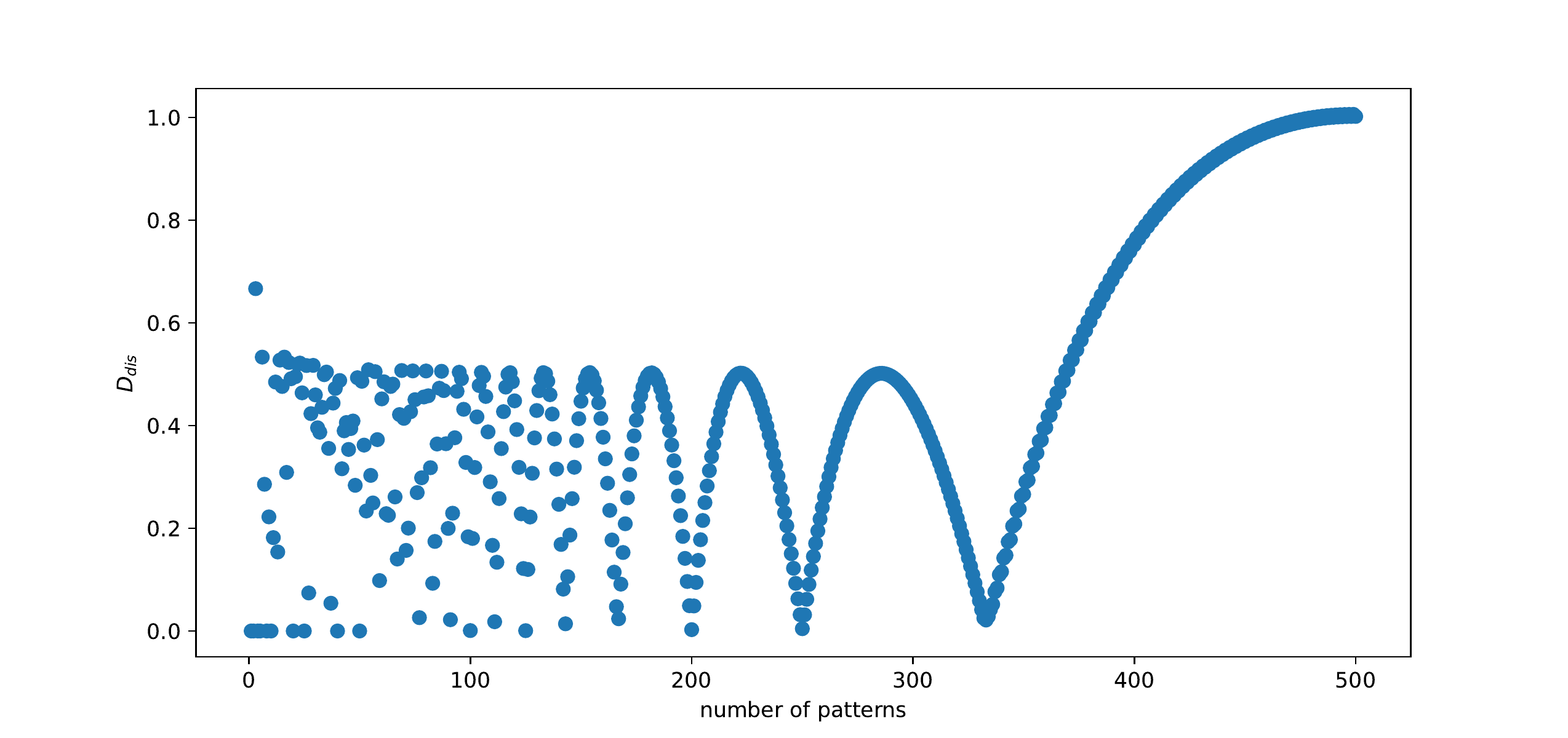}
    \caption{$D_\text{dis}$ of \eqref{eq:average_difference_size} for systems with 1000 states as a function of the number of patterns.}
    \label{fig:difference_pv}
\end{figure}

We consider the average difference $D_\text{dis}$ in the sizes of the basins
\begin{equation}
    D_\text{dis} = \frac{1}{k(k-1)/2}\sum_{i=1}^k\sum_{j=i+1}^k\big||B_i|-|B_j|\big|
    \label{eq:average_difference_size}
\end{equation}
and the performance is considered better for lower $D_\text{dis}$.
% In the plots we take the average of this average difference for the 100 disentangled systems mentioned above.\\
% The average difference in the sizes of the basins for systems with 5 patterns as a function of the number of states is plotted in Fig.~\ref{fig:difference_sv}. Starting from a number of states of 20 the general trend is that the average difference in the sizes of the basins increases linearly with the number of states. When the number of states becomes smaller than 20 it also increases. We also see from the values that the basins are really of a similar size. We see that for 60 states and 5 patterns, the basin size is $11 \pm 2$.  For 100 states and 5 patterns, the basin size is $19 \pm  2.6$.\\
Fig.~\ref{fig:difference_pv} shows the average of $D_\text{dis}$ for the 100 disentangled systems mentioned above for $N=1000$ states as a function of the number of patterns $k$. Note that when $N = km$ for some integer $m$, then $D_\text{dis}$ is minimal because most of the basins have the same size $m$. If not, the basins get different sizes because it is numerically just not possible for them all to have the same size.  In general, the performance remains really good.  
It takes approximately half a second for the disentanglement algorithm to finish on a standard PC for a tournament with $N=1000, k=10$.

% We see that it quickly decreases with an increasing number of patterns and then oscillates. An explanation for this is given in Appendix~\ref{section:oscillations}. We see that the average difference is typically low. The average difference in the sizes of the basins stays more or less the same at a value of a bit less than 2 for $N = 10-100$ where $N/k=10$. The average size of a basin is 9 in this case.\\ 

% \begin{figure}
%     \centering
%     \includegraphics[height=0.3\textheight]{sv_p5.pdf}
%     \caption{Average difference in the sizes of the basins for the disentanglement step for systems with 5 patterns as a function of the number of states.}
%     \label{fig:difference_sv}
% \end{figure}

% only mention
% In Fig.~\ref{fig:difference_sv_sdivp10} we show the average difference in the sizes of the basins as a function of the number of states where N/k=10. For the case where the number of states is 10 it is 0 because there is only one pattern.

% \begin{figure}
%     \centering
%     \includegraphics[height=0.3\textheight]{sv_px_sdivp10.pdf}
%     \caption{Average difference in the sizes of the basins as a function of the number of states where N/k=10.}
%     \label{fig:difference_sv_sdivp10}
% \end{figure}

\begin{figure}
    \centering
    \includegraphics[height=0.3\textheight]{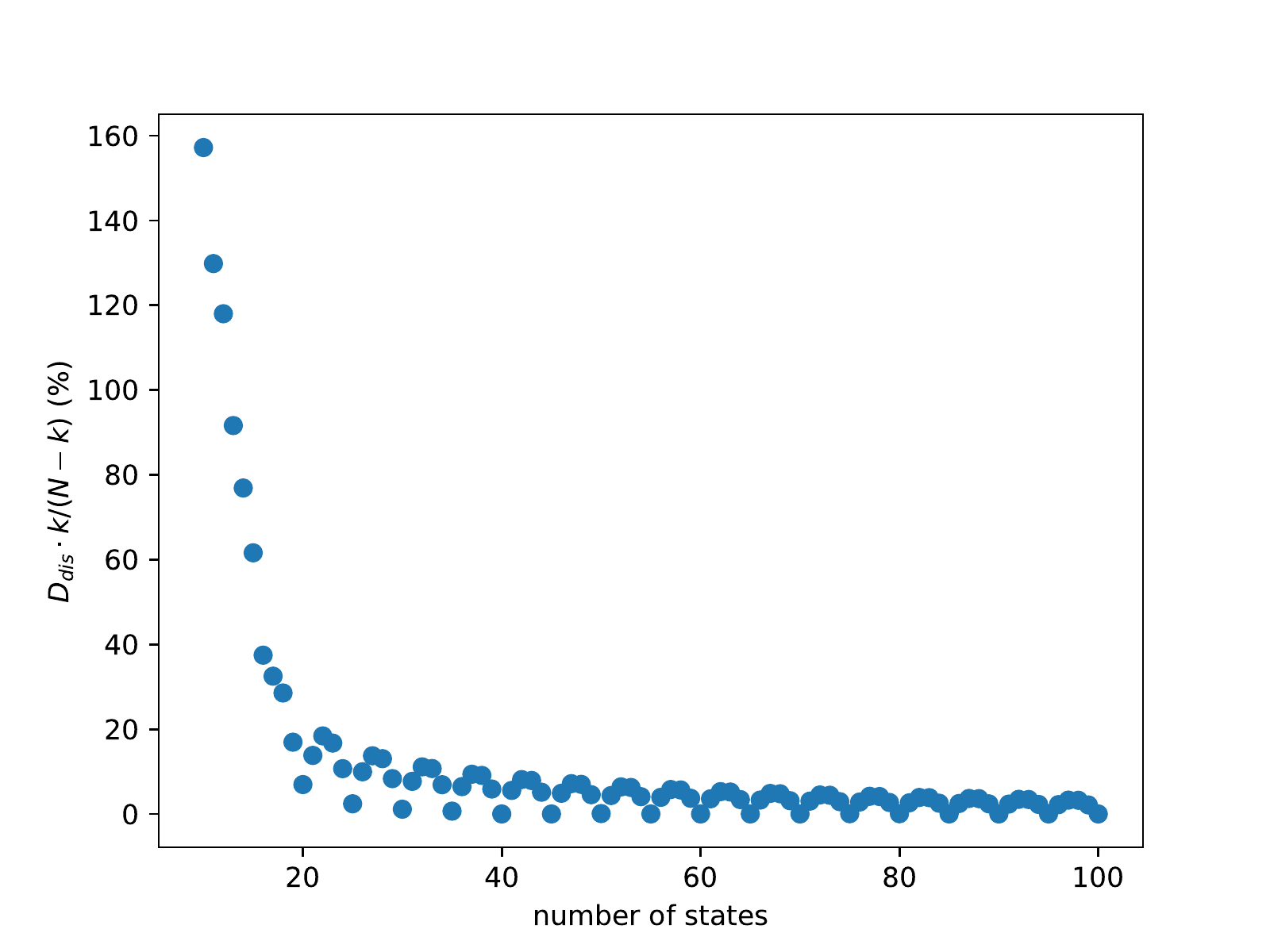}
    \caption{The relative  difference $D_\text{dis}\cdot k/(N-k)$ for systems with 5 patterns as a function of the number of states.}
    \label{fig:relative_difference_sv_p5}
\end{figure}

We also consider the relative average difference in the sizes of the basins, meaning to multiply  $D_\text{dis}$ with $k/(N-k)$. In Fig.~\ref{fig:relative_difference_sv_p5} we show this relative average difference in the sizes of the basins for systems with $k=5$ patterns as a function of the number $N$ of states. Again, we show the average of this average for the 100 disentangled systems. The minima are (again) for a number of states that is a multiple of the number of patterns. When the number of states gets smaller than 20 the relative average difference becomes larger because then some basins become empty and also the average size of the basins becomes smaller. For an increasing number of states, the relative average difference decreases because the average size of the basins increases.\\

%In general, the disentangling gets worse when $k/N$ gets larger.\\

The algorithm can be applied to any oriented graph but above, the performance was tested only for tournaments.  We next check when we randomly erase a fraction of the edges of the complete graph.\\
In Table \ref{table:incomplete_graph} the average over 100 disentangled systems of $D_\text{dis}$ is shown as a function of the fraction of arcs present for systems with 50 states and 5 patterns. Note that for this situation $10k=N$ and so $D_\text{dis}$ is minimal in case of a tournament. We conclude that good results are still possible, but as can be expected, $D_\text{dis}$ increases for stronger dilution.
\begin{table}[!h]
\begin{tabular}{r|cccccccc}
fraction of arcs kept & $0.3$ & $0.4$ & $0.5$ & $0.6$ & $0.7$ & $0.8$ & $0.9$ & $1$\\
\hline
$D_\text{dis}$ & 0.574 & 0.208 & 0.104 & 0.040 & 0.056 & 0.040 & 0.000 & 0.008\\
\end{tabular}
\caption{Difference $D_\text{dis}$ as a function of the fraction of arcs present for systems with 50 states and 5 patterns.}
\label{table:incomplete_graph}
\end{table}

\subsection{Performance of frenetic steering}\label{perf}
We start the walker from each state of the system and we look if it is in the (correct) pattern at time $T$ and stays long enough in the pattern. If that is the case, a success is recorded; otherwise a failure. We define the performance as the number of successes divided by the number of times we let the system evolve:  $P_\text{fren} = $ fraction of successes.\\

We generate 10 tournaments and for each of those,  we do the disentanglement. We thus obtain 10 disentangled systems. On each of those 10 disentangled systems, the frenetic steering is executed following the algorithm of Section \ref{sec:frenetic_steering}. We always show the average of $P_{\text{fren}}$ over these 10 random walks.\\ 
 
% \begin{figure}
%     \centering
%     \includegraphics[height=0.3\textheight]{s50_p10_av_n200_high.pdf}
%     \caption{Performance of the frenetic steering algorithm for systems with $50$ states, $10$ patterns and a training set size of $200$ as a function of $a$.}
%     \label{fig:s50_p10_av_n200_high}
% \end{figure}

As a first remark, increasing the driving value $\varepsilon$ always leads to better results; see Table~\ref{table:driving_value} where we show the performance  $P_{\text{fren}}$. That should not be surprising. 
\begin{table}[!h]
\begin{tabular}{r|cccccc}

$\exp(\ve)\,$ & $5$ & $6$ & $7$ & $8$ & $9$ & $10$\\
\hline
$P_\text{fren}$ \,& 0.662 & 0.694 & 0.752 & 0.776 & 0.818 & 0.844\\
\end{tabular}
\caption{$P_{\text{fren}}$ for a complete graph with $N=50$ states and 10 patterns for small values of $\varepsilon$. The value $a$ is given in \eqref{eq:a} and a training set size of 200 is taken.}
\label{table:driving_value}
\end{table}
In what follows we work with $\varepsilon = 5$ so that we have $a\simeq 1.64$ as initial activity parameter; see \eqref{eq:a}.\\

% \begin{figure}
%     \centering
%     \includegraphics[height=0.3\textheight]{s50_p10_a20_n200_ev.pdf}
%     \caption{Performance of the frenetic steering algorithm for systems with $50$ states, $10$ patterns, $a$ as in \ref{eq:a} and a training set size of $200$ as a function of $\varepsilon$, where $\log$ signifies the natural logarithm.}
%     \label{fig:s50_p10_a20_n200_ev}
% \end{figure}

%In Fig.~\ref{fig:s100_pv} we plot the performance for systems with 100 states as a function of the number of patterns. The performance keeps increasing with increasing number of patterns. We believe the main reason for this is that the basins become smaller when the number of patterns increases and that means there are less cycles found for each basin which means there are less arcs that have a pattern as start or end vertex and that makes the rate to leave a pattern smaller. An increasing number of patterns for a fixed number of states does not endanger recognition, on the contrary, it improves it. It is the desire that the basins should be of similar size that is endangered, as we have seen in the previous section. 
\begin{figure}
    \centering
    \subfigure[\,n=250]{
        \includegraphics[height=0.25\textheight]{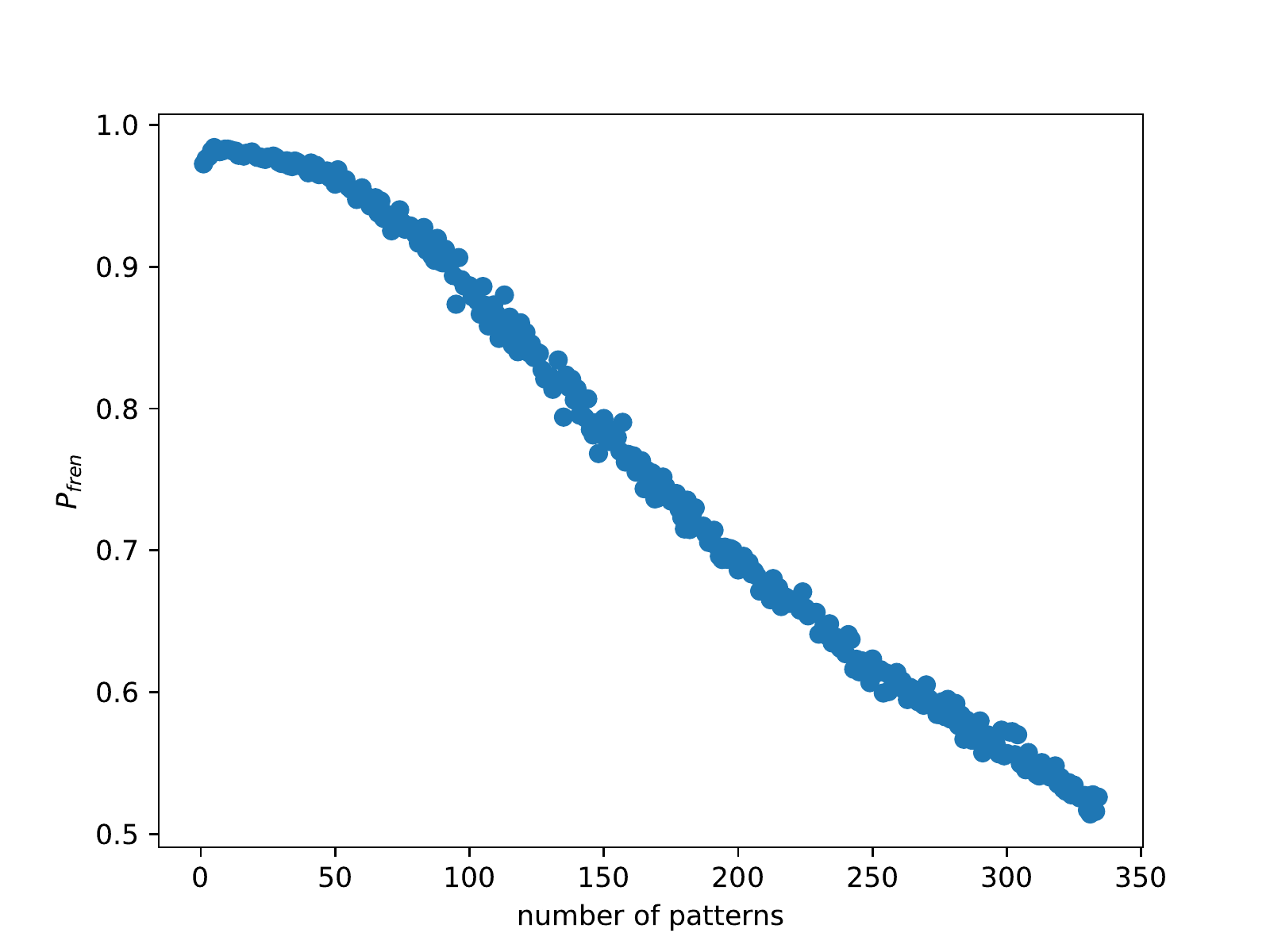}
        \label{fig:s1000_pv_n250}
    }
    \subfigure[\,n=500]{
        \includegraphics[height=0.25\textheight]{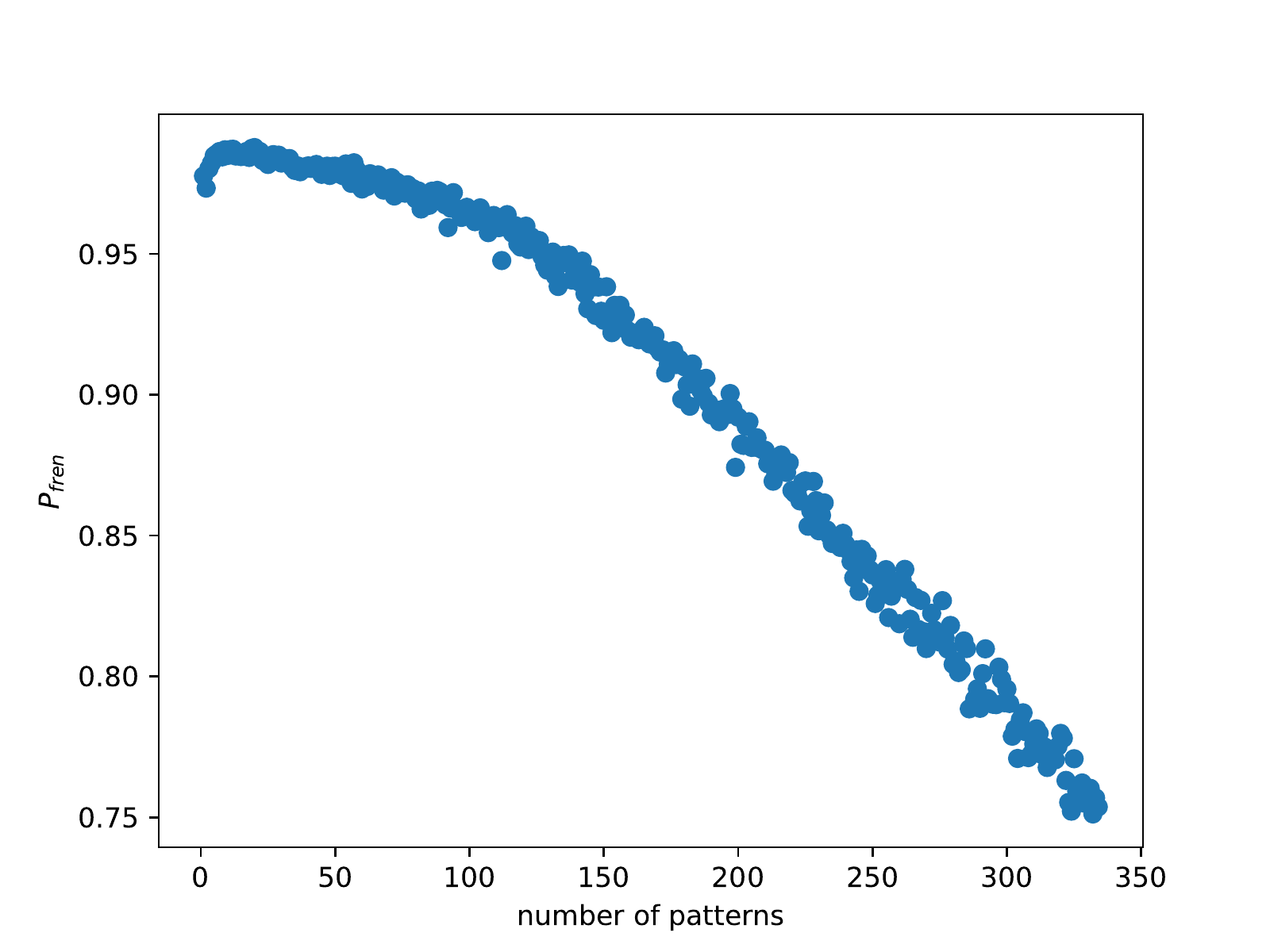} 
        \label{fig:s1000_pv_n500}
    }
    \subfigure[\,n=1000]{
        \includegraphics[height=0.25\textheight]{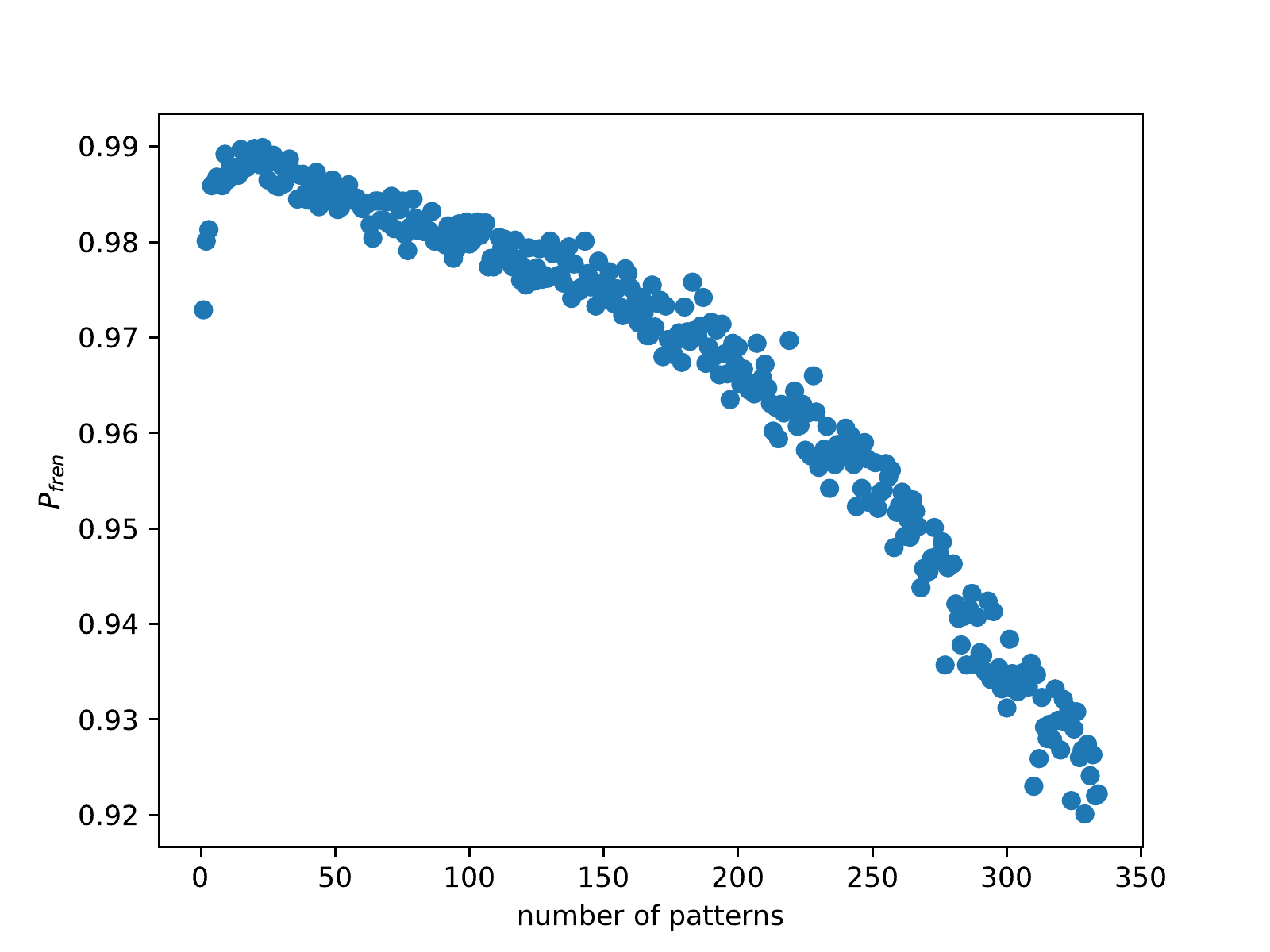}
        \label{fig:s1000_pv_n1000}
    }
    \subfigure[\,n=4000]{
        \includegraphics[height=0.25\textheight]{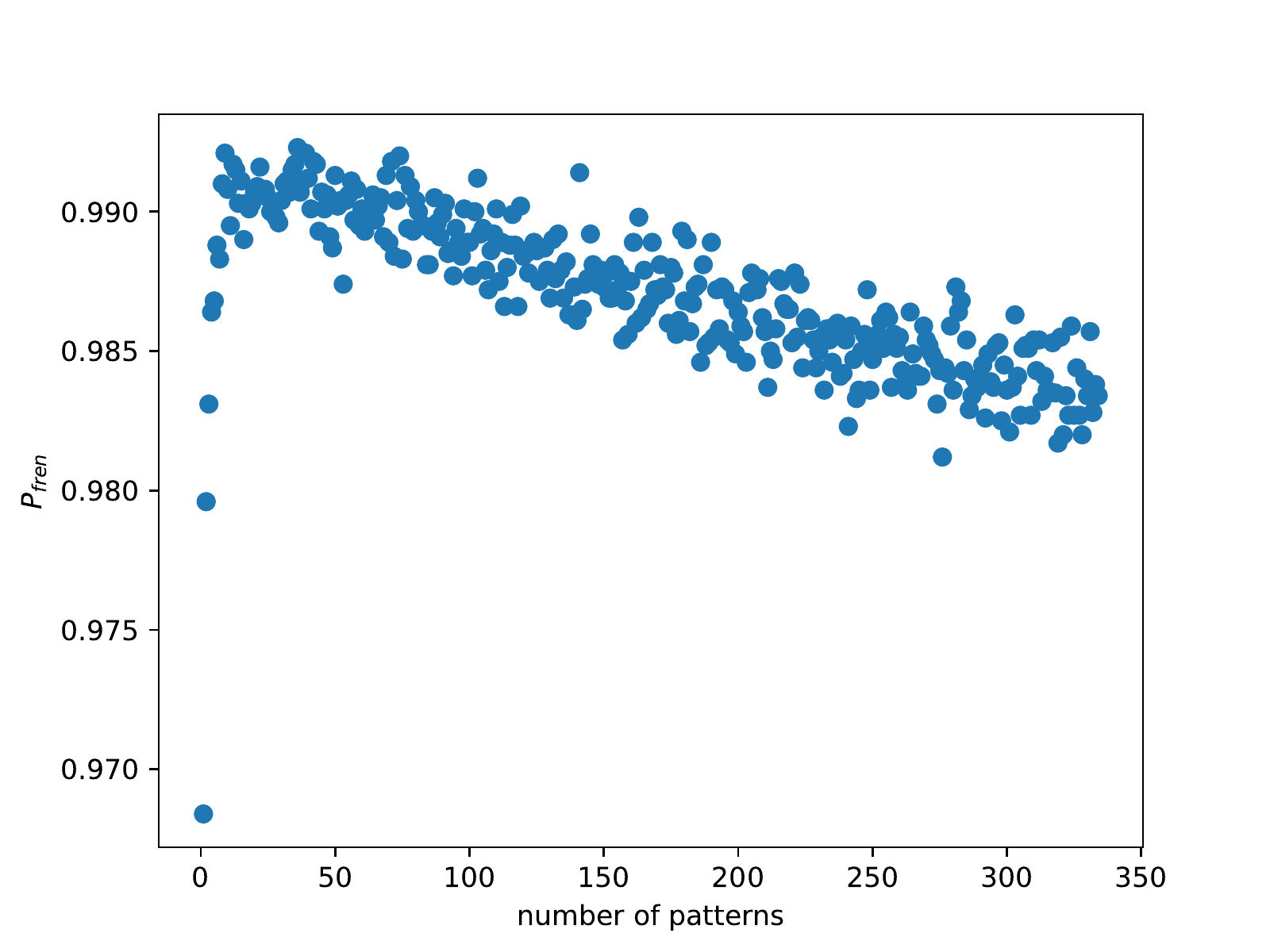}
        \label{fig:s1000_pv_n4000}
    }
    \caption{$P_{\text{fren}}$ for systems with $N=1000$ states as a function of the number of patterns for different values for the training set size $n$.}
    \label{fig:s1000_pv}
\end{figure}

In Fig.~\ref{fig:s1000_pv} we plot $P_{\text{fren}}$ for systems with 1000 states as a function of the number of patterns for different values of the training set size $n$. 
For every value of $n$, $P_{\text{fren}}$ decreases with an increasing number of patterns. That is how it should be and, in more detail, is probably due to the fact that for larger basins, the algorithm will more often  attempt to decrease the escape rate out of the pattern. Naturally, $P_{\text{fren}}$ increases for higher values of $n$; see Fig.~\ref{fig:s1000_pv}.\\ 

On a standard PC, the algorithm finishes in about 8 seconds for a disentangled system obtained from a tournament with 1000 states and 10 patterns.

\section{Conclusion}\label{con}
The morphology or physical chemistry of the brain is not leading to anything intelligent when passive.  Instead, in biological information--processing, we are dealing with nonequilibrium systems, well-outside the perturbative regime of close-to-equilibrium considerations.\\

In the present paper, we give up the idea that patterns are stored ``energetically,'' as parameters in the interaction potential between neurons.  We have introduced the concept of frenetic steering for the recall of a pattern.  In the first learning step, the disentangling models the nonlocal division of brain regions, where each component connects various neuronal conditions.  They are obtained starting from random but active wiring of the brain.  Once the regions get formed, the frenetic steering organizes the recall in each region, in the spirit of the Hebb rule where the synaptic strength between neighboring vertices or cells that are part of a signaling pathway is increased.
Indeed, the  frenetic steering is making changes to the time-symmetric activity parameters in the transition rates.  Those parameters give weights to connectivities and they are able to modify the occupation statistics exactly because of the (random) background driving.\\

\noindent {\bf Acknowledgment}:  Part of this work was started by Victor Kermans, \cite{kermans}, during his Master thesis with CM. No funding was received to assist with the preparation of this manuscript.\\

\section{Compliance with Ethical Standards}
No conflict of interest.\\


\begin{thebibliography}{triope}
\bibitem{sak1}
D.A.R.~Sakthivadivel. Characterizing the non-equilibrium dynamics of a neural cell.  arXiv:2102.09146v1 [q-bio.NC].

%\bibitem{sak2}
%D. A. R. Sakthivadivel. Formalising the use of the activation function in neural inference, 2021b.

\bibitem{ly}
C.~W.~Lynn, E.~J.~Cornblath, L.~Papadopoulos, M.~A.~Bertolero, and D.~S.~Bassett, Broken detailed balance and entropy production in the human brain. PNAS {\bf 118} (47) e2109889118,
(2021).

\bibitem{schr}
E.~ Schr\"odinger, What is Life? Cambridge University Press, 20th printing 2017 edition, 1967. %ISBN 978-1-107-60466-7. 

\bibitem{kar1}
J.~Karbowski, Metabolic constraints on synaptic learning and memory. \textit{Journal of Neurophysiology}, 122(4):1473-1490, 2019. %doi: 10.1152/jn.00092.2019. URL 
%\verb!https://doi.org/10.1152/jn.00092.2019!

\bibitem{kar2}
J.~Karbowski, Energetics of stochastic bcm type synaptic plasticity and storing of accurate information. \textit{Journal of computational neuroscience}, 2021. %ISSN 0929-5313.

\bibitem{al}
I.~Allaman and P.~J.~Magistretti. Chapter 12 - brain energy metabolism. In L. R. Squire, D. Berg, F. E. Bloom, S. du Lac, A. Ghosh, and N. C. Spitzer, editors, \textit{Fundamental Neuroscience (Fourth Edition)}, pages 261-284. Academic Press, San Diego, fourth edition edition, 2013. %ISBN 978-0-12-385870-2. doi: https://doi.org/10.1016/B978-0-12-385870-2.00012-3. URL 
%\verb!https://www.sciencedirect.com/science/article/pii/B9780123858702000123.!

\bibitem{swan}
L.~W.~Swanson. Chapter 2 - basic plan of the nervous system. In L. R. Squire, D. Berg, F. E. Bloom, S. du Lac, A. Ghosh, and N. C. Spitzer, editors, \textit{Fundamental Neuroscience (Fourth Edition)}, page 36. Academic Press, San Diego, fourth edition edition, 2013. %ISBN 978-0-12-385870-2. doi: https://doi.org/10.1016/B978-0-12-385870-2.00002-0. URL 
%\verb!https://www.sciencedirect.com/science/article/pii/B9780123858702000020!. 

%\bibitem{irr2}
%Christopher W. Lynn, Eli J. Cornblath, Lia Papadopoulos, and Danielle S. Bassett, Broken detailed balance and entropy %production in the human brain. \textit{PNAS} 118, November 2021. 

\bibitem{deco}
G.~Deco, Y.~Sanz Perl, J.~D.~Sitt, E.~Tagliazucchi, M.~L.~Kringelbach,
Deep learning the arrow of time in brain activity: characterising brain-environment behavioural interactions in health and disease.
bioRxiv 2021.07.02.450899.

\bibitem{seif}
A.~Seif, M.~Hafezi, and C.~Jarzynski,  Machine learning the thermodynamic arrow of time.
\textit{Nature Physics} 17, 105--113 (2021).

\bibitem{irr1}
L.~de la Fuente, F.~Zamberlan, H.~Bocaccio, M.~Kringelbach, G.~Deco, Y.~Sanz Perl, and E.~Tagliazucchi, Temporal irreversibility of neural dynamics as a signature of consciousness.
\textit{Cerebral Cortex}.  %doi: https://doi.org/10.1101/2021.09.02.458802 

\bibitem{shul}
R.~G.~Shulman and D.~L.~Rothman. Interpreting functional imaging studies in terms of neurotransmitter cycling. \textit{Proceedings of the National Academy of Sciences}, 95(20): 11993-11998, 1998. %ISSN 0027-8424. doi: 10.1073/pnas.95.20.11993. %URL 
%\verb!https://www.pnas.org/content/95/20/11993!

\bibitem{rai}
M.~E.~Raichle and M.~A.~Mintun. Brain work and brain imaging. \textit{Annual Review of Neuroscience}, 29(1):449-476, 2006. 
%doi: 10.1146/annurev.neuro.29.051605.112819. URL 
%\verb!https://doi.org/10.1146/annurev.neuro.29.051605.112819.!

\bibitem{fren}
C.~Maes, Frenesy: Time-symmetric dynamical activity in nonequilibria.  \textit{Physics Reports} 850, 1-33 (2020).

\bibitem{bai}
M.~Baiesi and C.~Maes, Life efficiency does not always increase with the dissipation rate. \textit{Journal of Physics Communications 2}, 045017 (2018).

\bibitem{nondis}
C.~Maes, Non-Dissipative Effects in Nonequilibrium Systems. \textit{SpringerBriefs in Complexity},  2018. %ISBN 978-3-319-67780-4 (2018). 

\bibitem{hopf}
J.~J.~Hopfield, Kinetic Proofreading: A New Mechanism for Reducing Errors in Biosynthetic Processes Requiring High Specificity.
Proc. Nat. Acad. Sci. {\bf 71}, 4135--4139 (1974).

\bibitem{cur}
C.~Maes, What decides the direction of a current?  Mathematics and Mechanics of Complex Systems {\bf 3--4}, 275--295 (2016).

\bibitem{heatb}
C.~Maes and K.~Neto\v{c}n\'{y}, {\it Heat bounds and the blowtorch theorem}. Annales Henri Poincar\`e \textbf{14}(5), 1193--1202 (2013).

\bibitem{lowT} 
C.~Maes, K.~Neto\v{c}n\'{y} and W. O'Kelly de Galway, {\it Low temperature behavior of nonequilibrium multilevel systems}. Journal of Physics A: Math. Theor. \textbf{47}, 035002 (2014).

\bibitem{land} 
R. Landauer, {\it Inadequacy of entropy and entropy derivatives in characterizing the steady state}. Physical review A, Atomic, molecular and optical physics, \textbf{12}(2), 636–638, (1975).

 \bibitem{int}
F.~Khodabandehlou, C.~Maes, K.~Neto\v{c}n\'{y}, Trees and forests for nonequilibrium purposes: an introduction to graphical representations.
J. Stat. Phys. {\bf 189} (3), (2022).

\bibitem{heb}
D.O.~Hebb, {\it The Organization of Behavior}. New York: Wiley \& Sons, 1949

\bibitem{python}
G.~Van Rossum, G. and F.L.~Drake, Python 3 Reference Manual, Scotts Valley, CA: CreateSpace, 2009.

\bibitem{numpy}
C.R.~Harris, K.J.~Millman, S.J.~van der Walt, {\it et al}. Array programming with NumPy. \textit{Nature 585}, 357–362 (2020). %DOI: 10.1038/s41586-020-2649-2

\bibitem{matplotlib}
J.D.~Hunter, Matplotlib: A 2D Graphics Environment, \textit{Computing in Science \& Engineering}, vol. 9, no. 3, pp. 90-95, 2007

\bibitem{git}
{\tt https://github.com/bramlefebvre/frenetic\_steering}

\bibitem{b-j}
J.~Bang-Jensen and G.Z.~Gutin, Digraphs: Theory, Algorithms and Applications. \textit{Springer Monographs in Mathematics}. Springer London : Imprint: Springer, London, 2nd ed. 2009. edition, 2009. %ISBN 978-0-85729-041-0.

\bibitem{douglas}
D.~B.~West, Introduction to graph theory. Pearson Education, Inc., First Indian reprint, 2002.%, ISBN 81-7808-830-4

% \bibitem{manoussakis}
% Y. Manoussakis, A linear time algorithm for finding hamiltonian cycles in tournaments. \textit{Discrete Applied Mathematics}, 36(2):199-201, 1992. %ISSN 0166-218X. doi: https://doi.org/10.1016/0166-218X(92)90233-Z. URL 
%\verb!https://www.sciencedirect.com/science/article/pii/0166218X9290233Z.!

\bibitem{kermans}
V.~Kermans and C.~Maes, Towards nonequilibrium aspects for neural networks. KU Leuven. Faculteit Wetenschappen. 2021. \href{https://kuleuven.limo.libis.be/discovery/fulldisplay?docid=alma9992668496501488\&context=L\&vid=32KUL_KUL:KULeuven\&search_scope=All_Content\&tab=all_content_tab\&lang=en}{url}

% \bibitem{paths}
% P.~Hell and M.~Rosenfeld, The complexity of finding generalized paths in tournaments, \textit{J. Algorithms 4} (1982) 303-309. 





\end{thebibliography}
\end{document}